\documentclass[a4paper,11pt]{article}
\usepackage[T1]{fontenc}
\usepackage{graphicx}
\usepackage{geometry}
\usepackage{hyperref}	
\usepackage{eso-pic}
\usepackage{pdfpages} 	
\usepackage{amsmath}
\usepackage{amssymb}
\usepackage{slashed}
\usepackage{graphicx}
\usepackage{graphics}
\usepackage{epsfig}
\usepackage{color}
\usepackage{authblk}
\usepackage{setspace}

\newcommand{\B}[1]{{\color{black}#1}}
\usepackage{ulem}
\usepackage{hyperref}

\title{\Large The quest for axions and other new light particles
\vspace*{1cm}}

\author[1]{\footnotesize  Working group: K.~Baker}
\author[2]{G.~Cantatore}
\author[3]{S.~A.~Cetin}
\author[4]{M.~Davenport}
\author[5]{K.~Desch}
\author[6]{B.~D\"obrich}
\author[7]{H.~Gies}
\author[8]{I.~G.~Irastorza}
\author[9]{J.~Jaeckel}
\author[6]{A.~Lindner}
\author[10]{T.~Papaevangelou}
\author[11]{M.~Pivovaroff}
\author[12]{G.~Raffelt}
\author[12]{J.~Redondo}
\author[6]{A.~Ringwald}
\author[13]{Y.~Semertzidis}
\author[4]{A.~Siemko}
\author[14]{M.~Sulc}
\author[15]{A.~Upadhye}
\author[16]{K.~Zioutas}

\affil[1]{\footnotesize Physics Department, Yale University, New Haven, CT USA}
\affil[2]{University and INFN Trieste, via valerio 2, 34127 Trieste, Italy}
\affil[3]{Dogus University, Istanbul, Turkey}
\affil[4]{CERN, CH-1211 Geneva-23, Switzerland}
\affil[5]{Physikalisches Institut, University of Bonn, Bonn, Germany}
\affil[6]{DESY, Notkestra\ss e 85, D-22607 Hamburg, Germany}
\affil[7]{Universit\"at Jena and Helmholtz Institute Jena, D-07743 Jena, Germany}
\affil[8]{Laboratorio de F\'isica Nuclear y Astropart\'iculas, Universidad de Zaragoza, Zaragoza, Spain}
\affil[9]{Institut f\"ur theoretische Physik, Universit\"at Heidelberg, Philosophenweg 16, 69120 Heidelberg, Germany}
\affil[10]{IRFU, Centre d'\'Etudes Nuclaires de Saclay, Gif-sur-Yvette, France}
\affil[11]{Lawrence Livermore National Laboratory, P.O. Box 808, Livermore, CA 94551-0808, USA}
\affil[12]{Max-Planck-Institut f\"ur Physik, F\"ohringer Ring 6, D-80805 M\"unchen, Germany}
\affil[13]{Brookhaven National Lab, Physics Dept., Upton, NY 11973-5000, USA}
\affil[14]{Technical University of Liberec, 46117, Czech Republic}
\affil[15]{Argonne National Laboratory, 9700 S. Cass Ave. Lemont, IL 60439, USA}
\affil[16]{University of Patras, GR 26504 Patras, Greece}
\affil[ ]{\ }
\date{}
\begin{document}
\thispagestyle{empty}
\maketitle

\begin{abstract} 
\normalsize
Standard Model extensions often predict low-mass and very weakly interacting particles, 
such as the axion. 
A number of small-scale experiments at the intensity/precision frontier are actively searching for these elusive particles,
complementing searches for physics beyond 
the Standard Model at
colliders.
Whilst a next generation of experiments  
will give access to a huge unexplored parameter space, a 
discovery would have a tremendous impact on our understanding of fundamental physics.
\end{abstract}

%
%

\newpage

\section{\large  The low-energy frontier of particle physics}
\begin{figure}[t]
\begin{center}
\includegraphics[width=15cm]{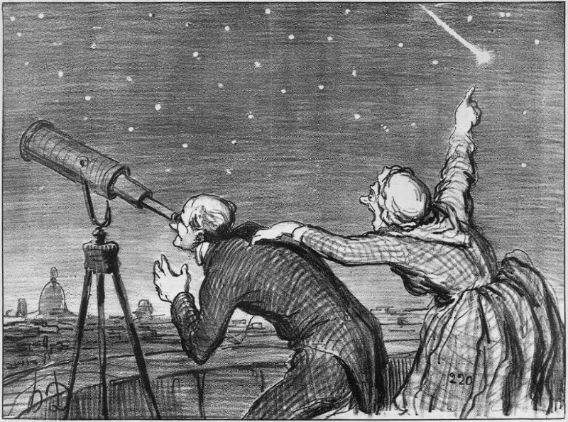}
\vspace{-0.2cm}
\caption{\footnotesize 
95\% of the universe are made of two mysterious substances, dark matter and dark energy that cannot be explained in the Standard Model.
By their very names it is clear that these things are somehow hidden from our view. New particles 
could hide by being very massive or by having extremely feeble interactions. 
It is clear that we need to look in all possible directions. In our quest for new physics high 
energy and low energy/high precision experiments nicely complement each other and together hopefully answer our questions to Nature.
{\small \copyright www. daumier-register.org}
\label{f:honore}
}
\end{center}
\end{figure}

The discovery of a Higgs boson at the LHC is an important step in the incredibly successful quest for
discovering the constituents of
the Standard Model of particle physics (SM). 
Yet, the SM is not a complete and fundamental theory since it does not explain the origin of Dark Matter and 
Dark Energy nor does it include gravity (see Fig.~\ref{f:honore}). 
Furthermore it does not provide satisfactory explanations for the values of its many parameters. 
It is therefore more timely than ever to look for physics beyond the SM. 
In this note 
we point out the importance of exploring the domain of very lightweight and very weakly coupling particles and discuss the
current and future experimental opportunities to do so.
Indeed, Weakly Interacting Slim (light) Particles (WISPs) are a feature of many well motivated theoretical models of 
fundamental physics. 
One prime example is the axion, a consequence of the Peccei-Quinn solution to the otherwise not explained problem of combined charge and parity conservation in quantum chromodynamics. 
Moreover, ultraviolet field or string theory embeddings of the SM often include hidden sectors of particles with extremely weak 
interactions with ordinary matter. 
The feeble strength of these interactions typically results from underlying new dynamics at energy scales much larger than
the electroweak scale. 
Thereby, probing these very weak interactions provides a new window to explore these high-energy scales and will give us 
crucial clues towards the understanding of the underlying structure of fundamental physics. 

At the same time (very) weak interactions seem to be a feature exhibited by the two most abundant yet also most mysterious 
substances in the universe,   Dark Matter and Dark Energy. Indeed axions and other WISPs are well-motivated Dark Matter candidates
and could
even open up a door to the understanding of Dark Energy.
Further motivation arises from anomalous white dwarf cooling and the propagation of very-high-energy gamma-rays in the intergalactic space. 

Experimental searches for WISPs require different approaches to those usually employed in particle physics. 
In particular, the very weak interactions involved demand experiments at the precision/intensity frontier. 
At low energies, intense lasers, strong magnetic fields, radio frequency technology as well as other techniques 
allow us to explore couplings many orders of magnitude smaller than those probed in collider experiments. 
Often, the techniques rely on coherence effects boosting experimental sensitivities.
A significant community has grown over the last 10 years and several experiments have already demonstrated the power 
of these approaches in relatively modest setups. 
Compared to collider experiments, this is still a very young field. 
Existing experiments can still be scaled-up to achieve sensitivity gains of many orders of magnitude and 
new experimental ideas arise very frequently with the potential of complementing existing schemes.

\section{\large Theory motivations for axions and other WISPs}\label{theory}

\subsection{\normalsize The strong CP problem and the QCD axion}
The combination of charge conjugation (C) and parity (P), CP, is violated in the weak interactions but it appears to be 
conserved in the electromagnetic and strong interactions. 
The latter constitutes a theoretical puzzle since the most general vacuum of quantum chromodynamics violates CP and, on top of that,    
the magnitude of this violation receives a contribution from the electroweak sector of 
the SM---which is well known to violate CP as well. From the experimental point of view,  
CP violation in quantum chromodynamics would manifest itself, e.g., in an electric dipole moment 
of the neutron.
The absence of any experimental evidence for an electric dipole moment 
of the neutron at the very strict level of $2.6\times10^{-26}\,\rm e\, cm$  is in
strong contradiction to the natural size of electric dipole moments expected from 
first principles which is at the $10^{-16}\,\rm e\, cm$ level. Thus
the experimental limit is at least ten orders of magnitude below the expected value. 

\begin{figure}[t]
\begin{center}
\includegraphics[width=12.5cm]{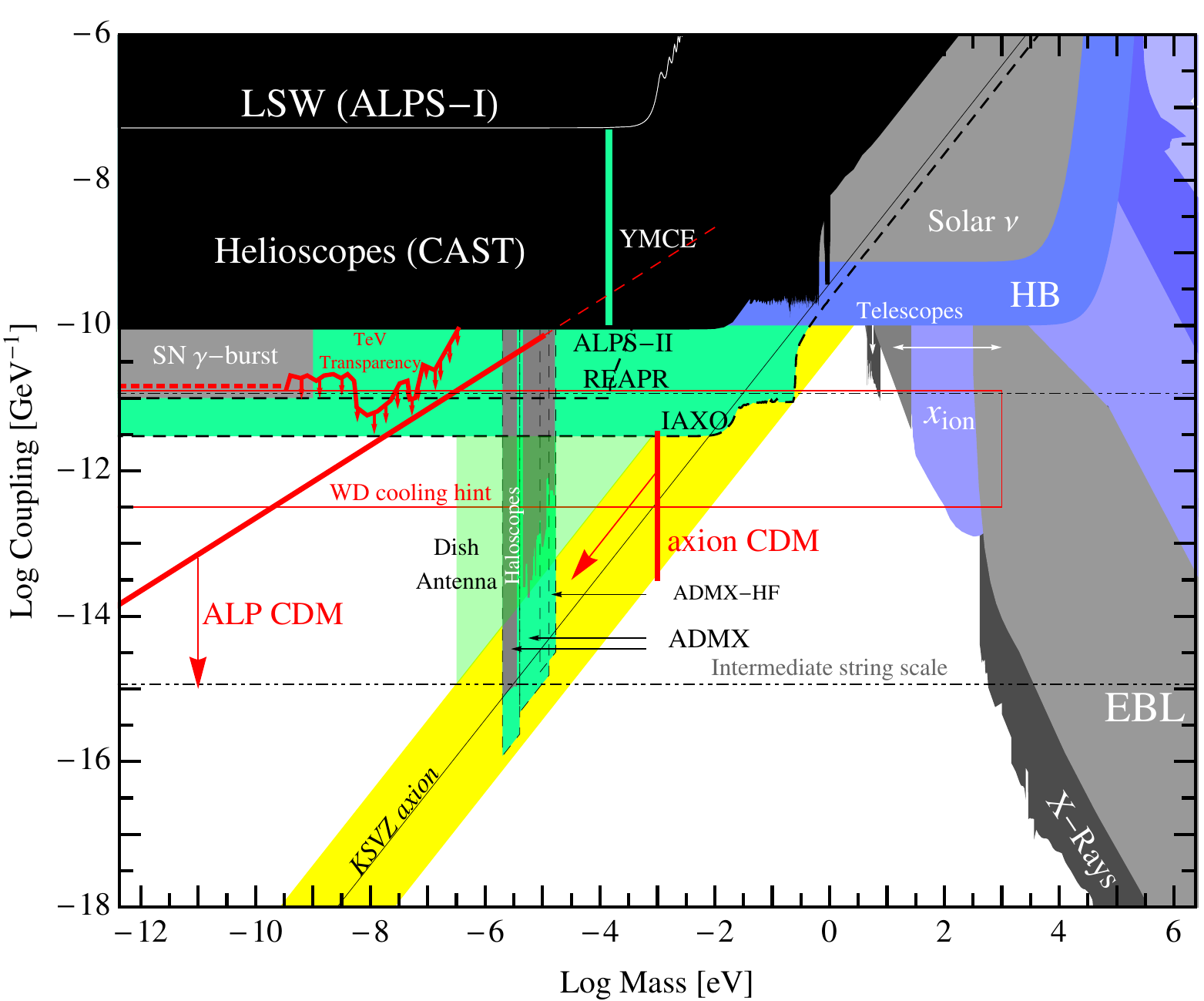}
\vspace{-0.2cm}
\caption{\footnotesize
Parameter space for axions and axion-like particles (see the text for explanations). 
Colored regions are: experimentally excluded regions (dark green), constraints from astronomical
observations (gray) or from astrophysical or cosmological arguments (blue), hints for axions and axion-like particles from astrophysics (red), and sensitivity of planned
experiments (light green). Figure adapted and updated from \cite{Hewett:2012ns}.
\label{f:ALPs}}
\end{center}
\end{figure}

Instead of relying on an accidental fine tuning of the CP violation parameters in QCD and weak interactions to explain this phenomenon,
Peccei and Quinn pointed out~\cite{Peccei:1977hh} in 1977 that the existence 
of a new symmetry, spontaneously broken at a high energy scale, denoted by $f_a$ in the following,
would automatically solve the strong CP problem if it is violated by the color anomaly. 
The axion is a new pseudo-scalar particle that arises as the pseudo-Nambu-Goldstone boson 
of this spontaneously broken symmetry~\cite{Wilczek:1977pj}. 
Discovering the axion is thus the way of assessing whether the Peccei Quinn mechanism is realized in nature. 
For most of the experimental searches, the key axion parameters are the mass $m_a$ and the anomalous coupling to two photons $g_{a\gamma}$. 
A summary of axion searches based on these two parameters is presented in Fig.~\ref{f:ALPs}.  
In this broad parameter space, axion models lie in a diagonal band (the ``axion band''\.) 
because the mass and couplings are inversely proportional to $f_a$ (which is unknown) and thus directly proportional to each other. 
This relation holds up to factors of order 1, because different realizations of the Peccei Quinn mechanism lead to slightly different
values of $g_{a\gamma}$.
  
Originally, it was assumed that $f_a$ is close to the weak scale $\sim 246$ GeV but collider experiments
quickly excluded this possibility, 
restricting $f_a$ to be larger than about $10^{5}$ GeV. Therefore, the axion must have very small mass, typically 
sub-eV, and feeble interactions. 
The axion is therefore a prototype example of a situation where underlying new physics at an extremely high energy 
scale leads to a new 
particle that can be probed in high precision low-energy experiments. 

\subsection{\normalsize WISPs in extensions of the Standard Model}
The case of the axion can be generalized to generic (pseudo-)scalars coupled to two photons, so-called axion-like particles (ALPs).
The relation between the mass and couplings of the axion is only intrinsic to the 
color anomaly of the Peccei Quinn symmetry, and thus generic ALPs can show up in all the parameter space of Fig.~\ref{f:ALPs}. 
 
Like the axion, ALPs can be realized as pseudo-Nambu-Goldstone bosons of symmetries broken at very high energies.  
Further motivation for their existence comes from string theory.  
One finds that these theories predict in general a rich spectrum of light (pseudo-)scalars with weak couplings. In particular, 
the compactification of type IIB string theory, where moduli-stabilization giving rise to  ALPs is well understood, can 
provide the axion plus many ALPs. 
Of particular interest are so-called intermediate string scales $M_s\sim 10^{10-11}$ GeV, as these can contribute to the 
natural explanation of several hierarchy problems in the SM. 
In these theories, ALPs generically exhibit a coupling to photons which lies in a 
range that is well accessible to some near-future WISP searches, see Fig.~\ref{f:ALPs}.

\begin{figure}[t]
\begin{center}
\includegraphics[width=15cm]{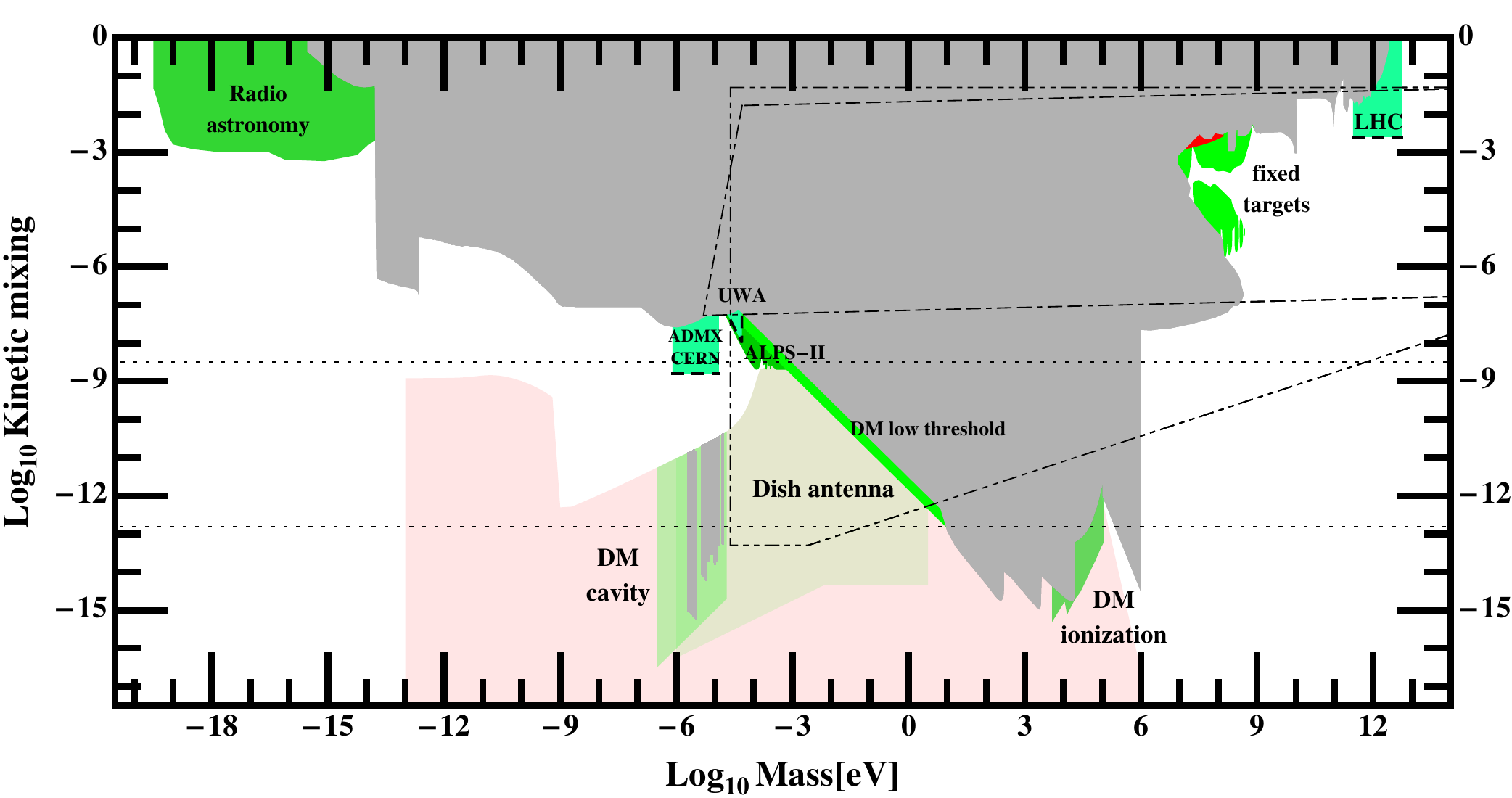}
\vspace{-0.2cm}
\caption{\footnotesize 
Parameter space for hidden photons (HPs).
The grey region indicates the parameter space already excluded by different experiments,
see \cite{Jaeckel:2013ija} for details. The reddish area in the bottom
indicated the region in which HPs could constitute Cold Dark Matter.
Various shades of green correspond to parameter regions accessible by planned and proposed
experiments.
\label{f:hps}
}
\end{center}
\end{figure}

Hidden sectors, i.e., particles with feeble interactions with ordinary matter, are a generic feature of field and string theory 
completions of the SM. 
For example, a ``hidden sector'' is commonly employed for supersymmetry breaking.
Hidden photons (HPs), i.e., gauge bosons of an extra U(1) gauge group, are natural ingredients of these hidden sectors.
Often, these hidden sectors interact with SM particles only through very heavy particles mediating between 
both sectors. 
They thus provide effective couplings -- kinetic mixing between photons and HPs
-- which makes them accessible to experimental searches, see \cite{Jaeckel:2013ija} and references therein.
In particular, 
if HPs have a non-zero rest-mass, kinetic mixing behaves as mass mixing and thus photon $\leftrightarrow$ HP 
oscillations occur (in analogy to oscillations among different neutrino flavors). 
This leads to the disappearance and regeneration of photons as they propagate in vacuum.
See Fig.~\ref{f:hps} for an overview of the HP parameter space, kinetic mixing vs. HP mass.
Moreover, the hidden sector naturally can also contain matter with fractional electric charge, often called 
minicharged particles (MCPs). 
Most prominently, they can emerge in theories which contain a hidden photon.
MCP searches provide an alternative observational window to the hidden sector and in particular can provide insight 
if the hidden photon turns out to be massless and thus not directly traceable.

Modifications of gravity proposed to explain the accelerated expansion of the universe often contain new scalar fields. 
If coupled to matter, they can very easily run into conflict with constraints on new long-range forces.
A certain class of these fields 
evades 
these bounds, e.g., by having either masses or couplings (or both) which depend on the ambient matter density. 
Nevertheless, their existence can still be probed in precision low-energy experiments and astrophysics.
A recent comprehensive review on WISPs and their detection can be found in~\cite{Jaeckel:2010ni,Ringwald:2012hr}. 

\section{\large Axions and WISPs in astrophysics and cosmology}\label{astrocosmo}

\subsection{\normalsize Dark matter}

Progress made in the last decade of cosmological observations has allowed us to establish a standard cosmological model. 
From the point of view of fundamental physics, its most remarkable feature is the need for at least 
two new ``substances'' not accounted for by the SM of particle physics: Dark Energy and   Dark Matter (DM). 
The former can be parametrized by a cosmological constant or some dynamical field and the latter by  non-relativistic 
particles of new species which interact feebly with SM particles and among themselves.

The particle interpretation of DM requires new fields beyond the SM. A popular example is 
the ``weakly interacting massive particle'' (WIMP) typically appearing in supersymmetric extensions of the SM. 
The prospects of exploring the  electroweak scale with the Large Hadron Collider has understandably focused the DM 
searches on the WIMP paradigm in the  last years, and the experimental community has devoted comparatively little effort 
to explore other possibilities. 

However, since the early 1980s axions are known to be very well motivated candidates for DM, see~\cite{Sikivie:2006ni} for a review.
Axion cold DM is produced non-thermally in the early universe by the vacuum-realignment mechanism 
and the decay of topological defects: axion strings and domain walls.
With the axion mass being so small, the phase-space density of cold DM axions is huge and it is 
indeed conceivable that they might form a Bose-Einstein condensate. 
This would produce peculiar structures in DM galactic halos.

Axions are not the only WISPs allowing for a solution to the Dark Matter question. 
The non-thermal production mechanisms attributed to axions are indeed  generic to bosonic WISPs such as ALPs and HPs. 
A wide range of WISP parameter space can generically contain models with adequate DM density.
Finally, it is noteworthy that axions and other WISPs are also produced thermally, i.e. in collisions of SM particles, 
thus contributing to some extent to either dark radiation or hot Dark Matter.

As the Large Hadron Collider at CERN has not yet observed any hint of supersymmetry, and after a decade of 
enormous progress in WIMP direct detection (more than 3 orders of magnitude improvement
in sensitivity in the WIMP-nucleon coupling cross-section) 
without a clear positive signal, the axion DM hypothesis stands out as increasingly interesting, and deserves serious attention. 
As reviewed in \cite{Ringwald:2012hr}, cosmological implications of the axion are well founded and represent a powerful motivation 
for experimental searches.

\subsection{\normalsize Dark energy}

The biggest puzzle in cosmology is Dark Energy, accounting for $\sim 70\%$ of the total energy density of the universe. 
Dynamical incarnations of Dark Energy are naturally related to the existence of very light and weakly interacting particles. 
While there are hardly any prospects to investigate Dark Energy directly in the laboratory, WISP searches could identify Dark Energy candidates. 
One such model, chameleon Dark Energy \cite{khourywelt}, is a canonical matter-coupled scalar field whose self-interaction implies a 
density-dependent effective mass. Thus, even in seemingly low-density environments such as the solar atmosphere or even the solar 
system, chameleons can gain a large effective mass that suppresses the chameleon-mediated fifth force between macroscopic bodies on observable length scales. 
Residual fifth forces, as well as particles of the chameleon field, may still be accessible to experiments.

\subsection{\normalsize Hints for WISPs from astrophysics}

A major astroparticle puzzle that has gained much attention over the past years is the observation of 
distant very high energetic gamma-ray sources with energies  above  $\mathcal{O}(100)$GeV.
Even if no absorbing matter blocks the way, gamma-ray absorption is to be expected as the gamma rays deplete through
electron-positron pair production when interacting with extragalactic background light. 
However, the observed energy spectra seem to lack the expected absorption features.
Axion-like particles can provide a resolution to this puzzle. 
Here, the anomalous transparency can be explained if photons convert into ALPs, travel unimpeded a  
fraction of the distance to us, and then 
reappear close to the solar neighborhood. 
Efficient photon-ALP conversions can happen in large scale astrophysical magnetic fields. 
The required range of masses and couplings is labeled ``transparency hint" in Fig.~\ref{f:ALPs}.
A second longstanding puzzle is the anomalously large cooling rate of white dwarfs suggested by the most recent 
studies of the white dwarf 
luminosity function
and the decrease of the pulsation period of the ZZ-Ceti G117-B15A.
The observed excess can be attributed to axions with $f_a\sim 10^{8-9}$ 
GeV if they couple to electrons. In a generic case the axion 
(or ALP) will also have a photon coupling. In Fig.~\ref{f:ALPs}, we have encircled in red the corresponding target region, 
labeled ``white dwarf (WD) cooling hint'', see \cite{Ringwald:2012hr,Hewett:2012ns} for references.
The existence of WISPs could manifest itself also in other dedicated astrophysical studies like spectral features related to strong magnetic fields. 
Such effects might deserve more attention in future.

\section{\large Experimental searches}\label{experiments}

Direct experimental searches for WISPs require a WISP source that can be astrophysical, cosmological or purely laboratory-based  \cite{Raffelt:1996wa,Jaeckel:2010ni}. 
Depending on the WISP under consideration, some sources are more relevant than others, although in general complementary 
studies are desirable. 
For axions, ALPs and HPs with masses above few meV, the sun and, \B{for HPs}, the early universe are the most copious
sources to consider. 
Helioscopes use the sun as a WISP source while haloscopes could detect Dark Matter composed of WISPs. Using stellar interiors and the early
universe as WISP sources requires trust in our ability to model these
extreme environments. It is therefore subject to astrophysical or cosmological uncertainties. 
Turning this argument inside out, the discovery of a  WISP could open up a window to study these extreme environments 
(as an example, WISPs could open up a new window in addition to neutrino physics and helioseismology to understand the interior of the sun). 
Needless to say, it is certainly advantageous to have complete control over the WISP source in a manner analogous to 
collider experiments. 
While at the moment experiments using astrophysical and cosmological sources are often more sensitive, it is likely that in the future
new techniques will make laboratory searches competitive or even superior in a variety of searches. Indeed 
low mass HPs, chameleons, galileons 
and certain classes of ALPs benefit from controlled vacuum conditions and they can be more easily 
produced in the laboratory, in particular by using intense laser fields.
Only laboratory experiments allow to measure the coupling strength of WISPs to photons for example, while helio- and haloscopes are sensitive 
to the product of flux and coupling strength only.

\subsection{\normalsize Helioscopes}

Helioscopes~\cite{Sikivie:1983ip} are designed to attempt the observation of the flux of WISPs generated in the sun 
and propagating to earth.
Different WISPs could be produced in the sun.
In principle, the flux of WISPs from the sun could reach up to 10\% of the solar photon
luminosity without the need for modifying current solar models. Note that for several WISPs,
experiments have already constrained this possibility.

The first axion helioscope search was carried out at Brookhaven National Lab in 1992 with a stationary
dipole magnet.
A second-generation experiment, the Tokyo axion helioscope (SUMICO), uses a more powerful
magnet and dynamic tracking of the sun. SUMICO has also performed
searches for 
solar HPs. For such searches, that do not rely on a magnetic field,
a dedicated Helioscope SHIPS (solar hidden photon search) experiment
exists in Hamburg.
A third-generation experiment, the CERN Axion
Solar Telescope (CAST), began data collection in 2003. It employs an LHC dipole test magnet
with an
elaborate elevation and azimuth drive to track the sun. CAST is also the first helioscope to use 
an X-ray optics
to focus the photon signal  as well as to apply some low background techniques from 
underground physics. Each generation of axion helioscope has achieved a six-fold 
improvement in sensitivity to the axion coupling constant over its predecessors, see Fig.~\ref{f:ALPs}. 

In addition to helioscopes, solar ALPs can be efficiently converted into photons in the high 
purity low background crystals used in some Dark Matter 
experiments
or in searches for neutrino-less double-beta decay. SOLAX, COSME, DAMA,
CDMS 
and 
Borexino 
have already presented
exclusion limits, which however are less stringent than CAST at low masses.
Larger and more sensitive detectors like GERDA and CUORE are planning to improve over previous results. 

To date, all axion helioscopes have recycled magnets built for other purposes. A significantly improved
sensitivity is to be expected from a custom-built magnet as with the proposed fourth-generation axion helioscope  
dubbed the International Axion Observatory (IAXO).
Beyond solar axions, IAXO has other potential physics cases. 
IAXO can be conceived as a unique generic facility for
axion/WISP research. By the operation of microwave cavities inside IAXO's magnet, IAXO could 
simultaneously search for solar and Dark Matter axions (see below). IAXO could also test whether 
solar processes can create HPs, chameleons and other WISPs.

\subsection{\normalsize Haloscopes}
Haloscope searches~\cite{Sikivie:1983ip}, originally proposed in the context of axion dark
matter, are based
on using the possible abundance of Dark Matter composed of WISPs all around us and
to exploit their coupling to ordinary photons which can then be detected.
As the DM particles are very
cold, their energy is
dominated by their mass.
Hence, in contrast to photons from solar WISPs, photons from DM WISPs can be considered as monochromatic.
Therefore the conversion of such WISPs into photons can be made more efficient by
employing a resonator having peak response at a frequency
corresponding to the energy of the produced photons, which is in turn equal to the 
energy of the incoming WISPs. 
The state of the art haloscope is the
axion Dark Matter search (ADMX)
which
uses a cryogenically cooled high-Q tunable microwave cavity immersed in an 8~T magnetic field. An additional project,
ADMX-HF, aiming at extending the range to higher frequencies has been 
recently approved at Yale University. Their current sensitivity and future reach are depicted in Fig.~\ref{f:ALPs}.

In addition to these experiments, further setups and ideas are being pursued to enhance the mass reach.
The Yale Microwave Cavity Experiment YMCE is a pioneer DM experiment at DM masses never 
explored before, but its planned sensitivity is for the moment 
not sufficient to reach the axion band (see Fig.~\ref{f:ALPs}). 
Also at DESY resonators and strong magnets are available to complement the mass reach of ADMX.
Preperatory studies ``WISPDMX'' in this direction have started. In addition it has been pointed out that
ALP and HP cold Dark Matter would excite antennas with a very narrow frequency signal that can be measured.   
Recently it has been suggested to use the fact that Dark Matter causes a time-varying
neutron electric dipole moment.
Experiments based on atom interferometry and storage ring electric dipole moments
which are currently under consideration could be much more sensitive to
oscillating electric dipole moments than to static ones.

%
\subsection{\normalsize Light shining through a wall}
For light shining through a wall experiments (LSW), the working principle is as follows:
LSW  is possible if incoming laser photons are converted into some kind of WISP
in front of a wall and reconverted into photons behind that wall (see~\cite{Redondo:2010dp} for a review). 
For some of these new particles \B{like ALPs}, the photon-to-WISP conversion processes 
(and vice versa) are induced by strong magnetic fields.
Current LSW experiments are not yet as sensitive as helio- and haloscopes, but offer full control
over the WISP production and do not rely on astrophysical or cosmological assumptions.
In addition, polarimetric measurements in magnetic fields, although typically less sensitive to new physics than 
LSW, can perform unprecedented precision tests of QED
as well as having some sensitivity to a range of WISPs.

Light shining through a wall experiments have
been successfully performed by a number of collaborations
at Toulouse (BMV),
CERN (OSQAR 
and microwave cavities),
Fermilab (GammeV), Jefferson Lab (LIPPS),
and DESY Hamburg (ALPS-I).
Future upgrades, such as ALPS-II 
and REAPR 
are laid
out to surpass current helioscope sensitivities.
LSW can also be done with ``light'' in the microwave regime
and first experiments in this direction have been performed.

\section{\large Rewards of the low-energy frontier}
Low-energy experiments enable us to explore for new particles in a complementary way to 
accelerator based experiments.
Whereas the latter allow us to look for new particles with high masses and moderate to large couplings, 
the former are particularly sensitive to particles with very small couplings and small masses.

As we have seen there are many convincing motivations for new light particles. They arise naturally in many 
extensions of the SM and they also allow for explanations of observations that are not accounted for
within the SM (important examples being the absence of CP violation in the strong interactions
and the nature of Dark Matter).

The present experimental work searching for particle candidates from the dark sector of the universe 
is by its very nature highly interdisciplinary, bringing together experts from various fields in physics. 
The cross-fertilization between
accelerator and particle detector technologies, lasers, quantum measurement science, nanotechnology,
etc., demonstrates the interdisciplinarity at its best, and new experimental ideas are 
suggested and implemented increasingly often.

To conclude: although significant challenges remain, the rewards of exploring the low-energy frontier of particle 
physics could be enormous.
Looking through the low-energy window may reveal fundamental insights on the underlying 
structure of nature and shed light on such important
mysteries as the origin of Dark Matter and Dark Energy.

\begin{spacing}{0.2}

\end{spacing}


\begin{thebibliography}{99}

\footnotesize

\bibitem{Peccei:1977hh} 
  R.~D.~Peccei and H.~R.~Quinn,
  Phys.\ Rev.\ Lett.\  {\bf 38}, 1440 (1977).
  %
\bibitem{Wilczek:1977pj} 
  F.~Wilczek,
  Phys.\ Rev.\ Lett.\  {\bf 40}, 279 (1978);
%
  S.~Weinberg,
  Phys.\ Rev.\ Lett.\  {\bf 40}, 223 (1978).
%

\bibitem{Hewett:2012ns} 
  J.~L.~Hewett {\it et al.},
  arXiv:1205.2671 [hep-ex].



\bibitem{Jaeckel:2013ija} 
  J.~Jaeckel,
  arXiv:1303.1821 [hep-ph].
  
\bibitem{Jaeckel:2010ni}
  J.~Jaeckel and A.~Ringwald,
  Ann.\ Rev.\ Nucl.\ Part.\ Sci.\  {\bf 60} (2010) 405. 

%


\bibitem{Ringwald:2012hr} 
  A.~Ringwald,
  Phys.\ Dark Univ.\  {\bf 1}, 116 (2012)
  [arXiv:1210.5081 [hep-ph]].
  
\bibitem{Sikivie:2006ni}
  P.~Sikivie,
  Lect.\ Notes Phys.\  {\bf 741}  19 (2008)
  [astro-ph/0610440].
  
  

  %
  \bibitem{khourywelt} 
  J. ~Khoury,  A. ~Weltman, 
  Phys.\ Rev. {\bf D69} 044026 (2004).
%
  
\bibitem{Raffelt:1996wa} 
  G.~G.~Raffelt,
  ``Stars as laboratories for fundamental physics: The astrophysics of neutrinos, axions, and other weakly interacting particles,''
  Chicago, USA: Univ. Pr. (1996) 664 p
%
%
%

%

\bibitem{Sikivie:1983ip}
  P.~Sikivie,
  Phys.\ Rev.\ Lett.\  {\bf 51} 1415 (1983) 
   [Erratum-ibid.\  {\bf 52} 695 (1984)].
%

%
\bibitem{Redondo:2010dp}
  J.~Redondo and A.~Ringwald,
  Contemp.\ Phys.\  {\bf 52} (2011) 211
  [arXiv:1011.3741 [hep-ph]].
  


\end{thebibliography}
\end{document}